# Two-qubit Gate of Combined Single Spin Rotation and Inter-dot Spin Exchange in a Double Quantum Dot

R. Brunner[1,2], Y.-S. Shin[1], T. Obata[1,3], M. Pioro-Ladrière[4], T. Kubo[5], K. Yoshida[1], T. Taniyama[6,7], Y. Tokura[1,5], S. Tarucha[1,3]

[1]Quantum Spin Information Project, ICORP, Japan Science and Technology Agency, Atsugi-shi, Kanagawa, 243-0198, Japan
[2] Institute of Physics, Montanuniversitaet Leoben, 8700, Austria
[3]Department of Applied Physics, University of Tokyo, Hongo, Bunkyo-ku, Tokyo, 113-8656, Japan
[4]Département de Physique, Université de Sherbrooke, Sherbrooke, Québec, J1K-2R1, Canada
[5]NTT Basic Research Laboratories, NTT Corporation, Atsugi-shi, Kanagawa, 243-0198, Japan
[6]Materials and Structures Laboratory, Tokyo Institute of Technology, 4259 Nagatsuta, Yokohama, 226-8503, Japan
[7]PRESTO, Japan Science and Technology Agency, 4-1-8 Honcho Kawaguchi, Saitama 332-0012, Japan

*Abstract*

A crucial requirement for quantum information processing is the realization of multiple-qubit quantum gates. Here, we demonstrate an electron spin based all-electrical two-qubit gate consisting of single spin rotations and inter-dot spin exchange in a double quantum dot. A partially entangled output state is obtained by the application of the two-qubit gate to an initial, uncorrelated state. We find that the degree of entanglement is controllable by the exchange operation time. The approach represents a key step towards the realization of universal multiple qubit gates.

In quantum information processing two-qubit gates have the ability to operate on basic algorithms including entanglement control, and therefore are essential to test, for example, a controlled NOT gate [1,2], the EPR paradox [3] or Bell inequalities [4]. Hence, their realization represents a major task in quantum information processing. Semiconductor quantum dots (QD) hailed for their potential scalability are outstanding candidates for solid state based quantum information processing [5]. Here, a single qubit, the smallest logical unit of a quantum circuit, is defined by the two spin states $|\uparrow\rangle$ and $|\downarrow\rangle$, respectively. Single spin control, crucial for the realization of single qubit gates has been demonstrated through magnetically [6], and electrically driven resonance (EDSR) [7-9]. However, two-qubit gates, act on four computational basis states denoted by $|\uparrow\rangle|\uparrow\rangle$, $|\uparrow\rangle|\downarrow\rangle$, $|\downarrow\rangle|\uparrow\rangle$, $|\downarrow\rangle|\downarrow\rangle$. The simplest two-qubit operation suitable to generate entanglement with spin qubits is a *"SWAP"* based on the exchange operation [1]. When the interaction between two qubits is turned "*on"* for a specific duration $\tau_{ex}$, that is $\tau_{ex} = \tau_{SWAP}$, the states $|\uparrow\rangle|\downarrow\rangle$ and $|\downarrow\rangle|\uparrow\rangle$ can be swapped to $|\downarrow\rangle|\uparrow\rangle$ and $|\uparrow\rangle|\downarrow\rangle$, respectively while $|\uparrow\rangle|\uparrow\rangle$ and $|\downarrow\rangle|\downarrow\rangle$ remain unchanged. A reduction of the operation time by a factor two, $\tau_{ex} = \tau_{SWAP}/2$, produces the $\sqrt{SWAP}$ or $SWAP^{n=1/2}$ gate, which has then the maximum entangling capability [10].

The electrical manipulation of exchange in a double QD has been demonstrated with a single singlet-triplet qubit [11]. However, the complete control of entanglement between two electron spins requires systematic manipulations of spin exchange and the possibility to address individual spins. Recently, an optical control of entanglement between two QD spins with a two-qubit gate has been achieved [12].

In this Letter we demonstrate an all-electrical two-qubit gate composed of single spin rotations and inter-dot spin exchange in a double QD with a novel split micro-magnet. The micro-magnet generates an inhomogeneous Zeeman field [7,8,13-15] necessary for the qubit operations. We show that (a) the two-qubit gate controls and probes the spin singlet component of the output state with a probability depending on the exchange operation time $\tau_{ex}$ and (b) the observed oscillations of the singlet probability with $\tau_{ex}$ strongly suggest the control of the degree of entanglement.

Fig. 1(a) shows the gate defined double QD with a split Cobalt (Co) micro-magnet. A quantum point contact (QPC) is used as a charge sensor [16] to map the charge stability diagram in Fig. 1(b). The charge state change is observed as a change in the QPC transconductance, $G_{QPC} = dI_{QPC}/dV_{PL}$ for the QPC current $I_{QPC}$ and the voltage $V_{PL}$ on the plunger-left (PL) gate. In the region of the stability diagram where $(N_L,N_R)$=(1,1) the double QD contains only two electrons, spatially separate from each other, one in each QD. Here, $N_L$ and $N_R$ is the number of electrons, for the left and right QD, respectively. Single spin rotations and inter-dot spin exchange manipulation are performed in the (1,1) region, along the detuning lines A, B and/or C under an external in-plane magnetic field $B_0$.

To rotate each electron spin of the double QD we use EDSR [7-9,13,15,17]. When the micro-magnet on top of the double QD is magnetized, well above saturation ($B_0 > 0.5$ T), along the z direction ($M_{Co}$), a stray magnetic field at the QD is generated. The stray field is composed of a slanted out-of-plane component $B_y(z)$ ( $dB_y/dz \sim T/\mu m$ ) and an inhomogeneous in-plane component $B_{in-plane}(x)$ ($<<B_0$) resulting in the Zeeman offset $\delta E_Z = E_{zL} - E_{zR}$ across the two QDs. We spatially displace with electric fields the electrons in the presence of $B_y(z)$ by applying microwaves (MW) to the top micro-magnet (Co gate). Single spin rotations occur when the MW frequency $f_{a.c.}$ matches the local Zeeman field $E_{zv=L,R}$ of the left or right QD. We set the QDs in the Pauli spin blockade (PSB) [18] and apply continuous (c.w.) MW at $f_{a.c.}$ by sweeping $B_0$ to measure two resonant peaks [Fig. 1(c)], one for spin rotations in the left QD, the other in the right QD [19]. PSB is established at an inter-dot energy detuning $\varepsilon$=0 at point A by the formation of the spin triplet state [T$_+$(1,1)=|↑⟩|↑⟩ or T$_-$(1,1)= |↓⟩|↓⟩] for $(N_L,N_R)$=(1,1) only when the Zeeman energy splitting

between the triplets $T_\pm(1,1)$ and $T_0(1,1)$ is larger than the fluctuating nuclear field (a few millitesla) [20]. For PSB due to spin selection rule $T_\pm(1,1)$ cannot change into the doubly occupied singlet S(0,2) with $(N_L,N_R)$=(0,2) and thereby current is blocked. However, EDSR can lift-off PSB with a spin rotation from $T_+(1,1)$ (or $T_-(1,1)$) to $|\downarrow\rangle|\uparrow\rangle$ or $|\uparrow\rangle|\downarrow\rangle$, followed by a transition to S(0,2). Note, that $T_0(1,1)$ is strongly hybridized to the singlet S(1,1) state by the Zeeman field gradient, and so is not subject to the blockade effect [7,8].

The control of specific spin rotations around the $x$ axis with a rotation angle $\theta$, in the Bloch sphere, is presented by measuring Rabi oscillations for both spins. Therefore, we set $B_0$ at each *c.w.* EDSR peak with $f_{a.c.}$= 11.1 GHz: $B_{0L} = 2$ T and $B_{0R} = 1.985$ T for the left and right QDs, respectively. Further, we apply voltage pulses non-adiabatically to Co and PL gates to change $\varepsilon$ [21]. In particular, we switch between two operation stages A ($\varepsilon$ =0) and B ($\varepsilon \approx$277 µeV, $\varepsilon$>>0) [Fig. 1(d)]. At stage A, in the PSB the two-electron state is initialized to either $T_-(1,1)$ or $T_+(1,1)$. Here, finite inter-dot tunnel coupling $t$ is present. However, in stage B where the exchange energy is negligible we perform controlled spin rotations with a rotation angle $\theta$ by applying pulsed MW with a duration $\tau_{EDSR}$. Finally, the readout at stage A allows the left electron to tunnel to the right dot with the probability depending on the spin rotation angle. The cycle [Fig.1(d)] of A→B→A is repeated continuously and lift-off of PSB at a given cycle modifies the average charge seen by the QPC. The averaged QPC signal is thus proportional to the probability of having anti-parallel spins $|\downarrow\rangle|\uparrow\rangle$ or $|\uparrow\rangle|\downarrow\rangle$. In Fig. 2(a) we then detect the averaged QPC signal which oscillates as a function of $\tau_{EDSR}$. The oscillations reveal a linear scaling of the oscillation frequency upon the square root of the MW power $P_{MW}$ or driving *ac* electric field amplitude for the left and right spins [Fig. 2(b)], a characteristic feature of Rabi oscillations [8]. $f_{Rabi}$ is higher for the left QD and so is the state fidelity reflecting a larger field gradient and MW field [21].

Next we prepare a two-qubit gate comprising controlled left spin *x*-rotations and inter-dot spin exchange between the QDs as illustrated in Fig. 3(a). We choose specific rotation angles for the left spin using pulsed MW at $B_{0L} = 2$ T. The inter-dot spin exchange operation is operated by manipulating the inter-dot exchange energy $J_0$ [1]. $J_0$ is defined as energy difference between the singlet S(1,1) and the triplet state $T_0$(1,1) and depends strongly on the relative energy detuning $\varepsilon$ of S(0,2) and S(1,1). It becomes large in the vicinity of zero detuning and vanishes for large detuning. To change $\varepsilon$ or $J_0$ we apply voltage pulses to PL and Co gates, establishing three quantum stages, namely A, B and C [Fig. 3(a)]. The operation starting at stage A either with $T_+$(1,1) or $T_-$(1,1) for $\varepsilon=0$ evolves by

$$T_{\pm}(1,1) \xrightarrow{L\frac{3\pi}{2}} \frac{|\uparrow\rangle \pm |\downarrow\rangle i}{\sqrt{2}} \otimes |\uparrow\rangle \xrightarrow{J_0:\tau_{ex}} |\psi_1\rangle \xrightarrow{L\frac{\pi}{2}} |\psi_2\rangle, \qquad (1)$$

where $L\frac{3\pi}{2}$, and $L\frac{\pi}{2}$ in stage B represents the specific $\frac{3\pi}{2}$, and $\frac{\pi}{2}$ rotations, respectively around the *x* axis. At stage B the inter-dot tunnelling and therefore $J_0$ is negligible for $\varepsilon \approx 277$ μeV. The quantum operation $J_0:\tau_{ex}$ at stage C represents the two-qubit exchange operation. Here, for $\varepsilon \to 0$, e.g. 27.70 μeV, the exchange is controlled by the operation time or hold time $\tau_{ex}$. $|\psi_1\rangle$ is then the two-qubit state after the controlled rotation $L\frac{3\pi}{2}$ and exchange operation. After $L\frac{\pi}{2}$, $|\psi_1\rangle$ is finally transformed to the output state $|\psi_2\rangle$. Note, that the state fidelity of the two single spin rotations in stage B ($L\frac{3\pi}{2}$, and $L\frac{\pi}{2}$) strongly influences that of the presented two-qubit gate operation [21]. The cycle A through C is repeated continuously. Assuming an initialization to $T_+$(1,1) the wavefunction at the output controlled by $\tau_{ex}$ is e.g. $|\psi_2\rangle = T_+(1,1)$ for *no exchange operation* ( $NOP = SWAP^{n=0,2,4,\cdots}$ ), and $|\psi_2\rangle = \frac{1}{2}[T_+(1,1) + T_-(1,1) - \sqrt{2}iS(1,1)]$ for $SWAP^{n=1,3,5,\cdots}$. The single spin rotation angles are chosen such that

$|\psi_2\rangle$ has only $T_\pm$ and S components irrespective of the initial state ($T_+$ or $T_-$). Because of PSB the triplets $T_\pm(1,1)$ itself do not bring about the change of charge, only the singlet component of the output state gives rise to charge transitions (from (1,1) to (0,2)) at the readout stage [22]. The charge sensor readout is thereby a direct measurement of the probability $P_S = |\langle S|\psi_2\rangle|^2$. Therefore, in case of $SWAP^{n=1,3,5,\ldots}$, only $\sqrt{2}iS(1,1)$ is probed in $|\psi_2\rangle$. However, for $NOP$ no charge transfer is detected resulting in a minimum of the QPC signal. In Fig. 3(b) we plot the change of the charge state measured by the QPC as a function of $\tau_{ex}$ and detuning $\varepsilon$ or $J_0$. The measurement exhibits periodic oscillations as a function of both parameters. The experimental data agree well with a model calculation of $P_S$ [21]. The model includes the effect of finite $\delta E_Z$, and nuclear field fluctuations [15]. Maxima in Fig. 3(b) appear when the exchange operation is $SWAP^{n=1,3,5,\ldots}$ for $\tau_{ex} = (2k+1)\tau_{SWAP}$ and minima when $\tau_{ex} = k\tau_{NOP}$ with $\tau_{NOP} = 2\tau_{SWAP}$, where $k$=0, 1, 2,…. for $NOP$. $SWAP^{n=1/2}$ is obtained for $\tau_{ex} = \tau_{SWAP}/2$. That is, the two-qubit gate combined with PSB enables the control and detection of the singlet component in the output state with the finding probability depending on the exchange operation time $\tau_{ex}$. Using the model calculation allows us to extract the operation time $\tau_{SWAP}$ for $SWAP^{n=1}$, defined as half the oscillation period. In Fig. 4(a), we investigate the dependence of $\tau_{SWAP}^{-1}$ on $\varepsilon$. As expected, $\tau_{SWAP}$ is getting shorter with decreasing $t$ [1]. In addition the inset of Fig. 4(a) shows the effect of inter-dot tunnel coupling $t$ on $\tau_{SWAP}$. Note, the exchange energy depends on $\varepsilon$ and $t$, where $t \approx \sqrt{\frac{1}{2}J_0\varepsilon}$, for $\varepsilon \gg t > \delta E_Z$ [23,24]. The data points in Fig. 4(a) are only reproduced if we assume $\delta E_Z$ to be varying linearly with $\varepsilon$ [21]. $J_0$ defined by the oscillation period and $\delta E_Z$ obtained from the fit in Fig. 4(a) yields a ratio $\delta E_Z/J_0$ necessary for the calculated $P_S$ to resemble the experimental data in Fig. 3(b).

Finally, to evaluate the degree of entanglement between the two electron spins we calculate the concurrence $C$ [25] for the output state $|\psi_2\rangle$ [21] as a function of $\tau_{SWAP}$. For maximally entangled qubits $C(\tau_{ex})$=1, and uncorrelated qubits $C(\tau_{ex})$=0. The analytical expression of $C$ by neglecting nuclear spin fluctuations but including the effect of $\delta E_Z$ is given by:

$$C = \frac{|\sin\sqrt{1+\Delta^2}\alpha|}{1+\Delta^2} \times \sqrt{(1+\Delta^2)\cos^2\sqrt{1+\Delta^2}\alpha + \Delta^2\sin^2\sqrt{1+\Delta^2}\alpha} \qquad (2)$$

with $\Delta \equiv \delta E_Z/J_0$ and $\alpha \equiv J_0\tau_{ex}/2$ [21]. Fig. 4(b) shows the calculated $C$ as a function of $\tau_{ex}$ and $J_0$. $C$ of $|\psi_1\rangle$ [21] is zero at e.g. $\tau_{ex}$=0 and $\tau_{SWAP}$ or maximal ($C$=1/2) for $SWAP^{n=1/2}$ at $\tau_{ex} = \tau_{SWAP}/2$ when $\delta E_Z = 0$ [25]. When $\delta E_Z \neq 0$ the $\tau_{ex}$ dependence of $C$ is slightly modified by reducing the maximal value of $C$ [21]. However, the calculated $C$ in comparison with the observed $P_S$ gives evidence for the control of the degree of entanglement with $\tau_{ex}$.

We have demonstrated an all-electrical two-qubit gate comprised of controlled single spin rotations and spin exchange in a double quantum dot. Therefore, we used a micro-magnet to drive spin rotations under *ac* electric fields and voltage pulses to control the exchange interaction. The two-qubit gate generates a singlet component in the output state, which is probed directly by charge sensing. In addition, we calculated the degree of entanglement using the parameters derived from the experiment. Finally, we propose that with faster single spin rotations the two-qubit gate implemented here would be highly suitable to test in future experiments controlled NOT [1,2], the EPR paradox [3] or Bell inequalities [4].

ACKNOWLEDGEMENT

This work was supported financially by Grants-in-Aid for Scientific Research S (No.19104007), B (No. 18340081), and Project for Developing Innovation Systems of MEXT, IARPA project "Multi-Qubit Coherent Operations" through Harvard University, MEXT KAKENHI "Quantum Cybernetics" project and Funding Program for World-Leading Innovative R&D on Science and Technology（FIRST), the JSPS and CIFAR.

REFERENCES

[1] D. Loss and D. P. DiVincenzo, Phys. Rev. A **57**,120 (1998).

[2] M. A. Nielsen and I. L. Chuang, Cambridge University Press (2000).

[3] A. Einstein, B. Podolsky, N. Rosen, Phys. Rev. **47**, 777 (1935).

[4] J. S. Bell, Rev. Mod. Phys. **38**, 447 (1966).

[5] R. G. Clark, (ed.) Quant. Inform. Comput. 1 (special issue on implementation of quantum computation) 1-50 (2001).

[6] F. H. L. Koppens *et al*., Nature **442**, 766 (2006).

[7] M. Pioro-Ladrière *et al.,* Nat. Phys. **4**, 776 (2008).

[8] T. Obata *et al*., Phys. Rev. **B 81**, 085317 (2010).

[9] S. Nadj-Perge *et al*., Nature **468**, 1084 (2010).

[10] S. Balakrishnan and R. Sankaranarayanan, Phys. Rev. **A 78**, 052305 (2008).

[11] J. R. Petta, *et al.* Science **309**, 2180 (2005).

[12] D. Kim *et al*., Nat. Phys. **7**, 223 (2011).

[13] Y. Tokura *et al*., Phys. Rev. Lett. **96**, 047202 (2006).

[14] Y.-S. Shin *et al.* Phys. Rev. Lett. **104**, 046802 (2010).

[15] T. Obata et al. J. Phys.: Conf. Ser. **193**, 012046 (2009).

[16] M. Field *et al*., Phys. Rev. Lett. **70**, 1311 (1993).

[17] V. N. Golovach, M. Borhani, D. Loss, Phys. Rev. **B. 74**, 165319 (2006).

[18] K. Ono *et al*., Science **297**, 1313 (2002).

[19] The peak broadening which strongly depends on the $B_0$ sweep rate and sweep direction is caused by dynamical nuclear polarization which can fluctuate the $B_0$ condition of EDSR see [15].

[20] F. H. L. Koppens *et al*., Science **309**, 1346 (2005).

[21] See supplementary material chapter A1, A2, A3, A4 and figures S1, S2 and S3.

[22] Note, that PSB cannot be lifted by the population of excited states in the double QD, since the pulsing $\varepsilon$ range used here is small enough not to populate any other excited state. The excitation energy is on the order of meV significantly larger than $\varepsilon$ used here. This argument is also valid for the single-qubit gate.

[23] R. Hanson and G. Burkard, Phys. Rev. Lett. **98**, 050502 (2007).

[24] X. Hu and S. Das Shama, Phys. Rev**. A 68**, 052310 (2003).

[25] S. Hill and W. K. Wooters, Phys. Rev. Lett. **78**, 5022 (1997).

**FIGURE CAPTIONS**

FIG. 1 (a) Scanning electron microscopy image of the device fabricated on top of an AlGaAs/GaAs heterostructure showing the Ti/Au gates (light grey) and the split Cobalt (Co) magnet (yellow) separated from the gate contacts by a calixarene layer. Gates R (right) and L (left) to control $N_R$ and $N_L$, C (center) to control the inter-dot tunnel coupling $t$. Fast voltage pulses are applied to the Co and plunger-left (PL) gates. A MW voltage $V_{ac}$ is applied to the upper part of the magnet. $G_{QPC}$ is measured by modulating the PL gate voltage $V_{PL}$. (b) Stability diagram ($G_{QPC}$ v.s. $V_L$ and $V_R$ applied to the gates $L$, and $R$, respectively) in PSB regime, $B_0$=1 T (no MW). Source (S)-drain (D) bias is 1.5 mV. $\varepsilon$ is measured from the ($N_L$,$N_R$)=(0,2)-(1,1) boundary (dotted line: $\varepsilon$ =0) to the (1,1) ((0,2)) region. Dotted line highlights the experimentally obtained region where the lift-off of PSB at EDSR occurs. Schematically further detuning lines labeled B and C are shown. (c) c.w. EDSR for the left and right spin. PSB is lifted on resonance for the left (red) and right (blue) QD spin ($V_C$=-1.090 V, $f_{ac}$=5.6 GHz). EDSR peak separation: $\Delta B_0$ =15 ± 5 mT. The g-factor from $f_{a.c.}$ vs. $B_0$ : g=-0.394±0.001. (d) Measurement cycle for controlled single spin rotations with source (S), drain (D), left (L) and right (R) QD. Repetition period ~9 µs and repeated ~100 times

FIG. 2 (a) Rabi oscillations for the left (red) and right (blue) ($B_{0L}$ =2 T and $B_{0R}$ =1.985 T, $V_C$=-1.090 V, $f_{a.c.}$=11 GHz). $\delta G_{QPC}$ is the difference in $G_{QPC}$ between the on-resonance and off-resonance conditions with $B_0$ as a parameter. (b) Rabi oscillation frequency $f_{Rabi}$ as a function of the square root of MW power, $P_{MW}^{1/2}$, for the left (red) and right (blue) QD spin ($B_{0L}$ =2 T and $B_{0R}$ =1.985 T).

FIG. 3 (a) Cycle of the two-qubit gate operation with source (S), drain (D), left (L) and right (R) QDs. (b) Result of two-qubit measurement for $\varepsilon$ = 27.70 (A), 55.40 (B), 83.10 (C), 138.50 (D) ($V_C$=-1.0845 V, $f_{Rabi}$=1.2 MHz, $B_0$=2 T). Contour plot showing $J_0$ vs. $\tau_{ex}$ indicating $P_S$. We use the ratio $\delta E_Z/J_0$ as a fitting parameter to reproduce the experimental data and find that all data (A) to (D) measured for various detuning values are consistent with the calculation by taking $\delta E_Z/J_0 \approx 0.74$ ($SWAP^{n=1,3,5,\cdots}$: red, $NOP$: black). $f_{Rabi}$=1.2 MHz and nuclear spin variance for left and right spin is 0.275±0.025 MHz. Clear dependence on $\tau_{ex}$ and $\varepsilon$ is demonstrated with $\delta E_Z/J_0$ =0.69 (A), 0.73 (B), 0.78 (C), 0.77 (D) which gives on average 0.74. Yellow solid curves represent $P_S$ for (A) - (D) vs. $\tau_{ex}$. (Curves are offset for clarity).

FIG. 4 (a) $\tau_{SWAP}^{-1}$, as a function of $\varepsilon$ for different tunnel coupling. Blue: $V_C$ = -1.082 V ($t$=1.13± 0.1 μeV). Red: -1.0845 V ($t$ =0.98 ± 0.1 μeV). Solid curves: Fits for a linearly varying Zeeman gradient, $\Delta B_0 \equiv B_{0L} - B_{0R} = a + b\varepsilon$, with g = -0.4. Fitting parameters $a$ and $b$ are assumed to be independent of $V_C$ ($a$=-7.1 ± 0.4 mT, $b$=-24.4 ± 3 T/eV and $t$=1.13 ± 0.1 μeV ($t$=0.98 ± 0.1 μeV)). As expected the $V_C$ primarily controls $t$. Inset: $\tau_{SWAP}$ obtained for $\varepsilon$=27.70 ± 1.50 μeV *vs.* $t$ for $V_C$=-1.081, -1.082, -1.083, -1.0845, and -1.086 V; (from right to left), $\varepsilon$=*const*. The shortest $\tau_{SWAP}$ obtained here is ≈10 ns. (b) $C$ *vs.* $\tau_{ex}$ and $J_0$ for the average ratio $\delta E_Z/J_0$ =0.74 used for the $P_S$ calculation. $C = 50$ % for maximum entanglement at $\tau_{ex} = \tau_{SWAP}(2k+1)/2$, $k$=0,1,2,... .

# FIGURES

FIG.1 (color)

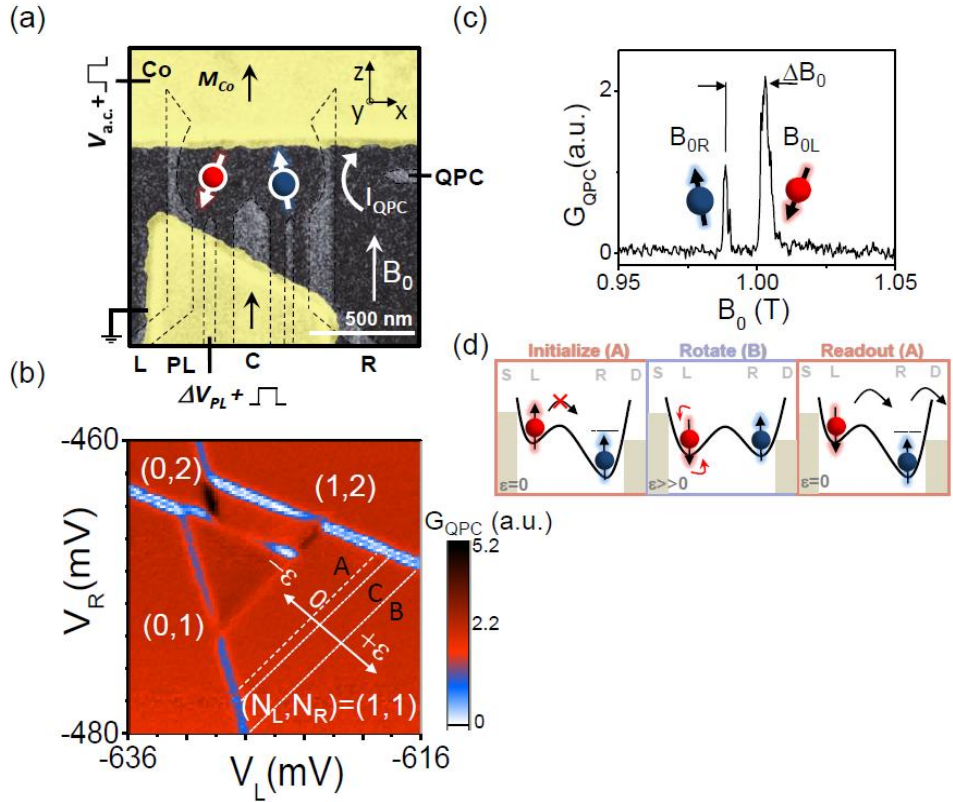

FIG.2 (color)

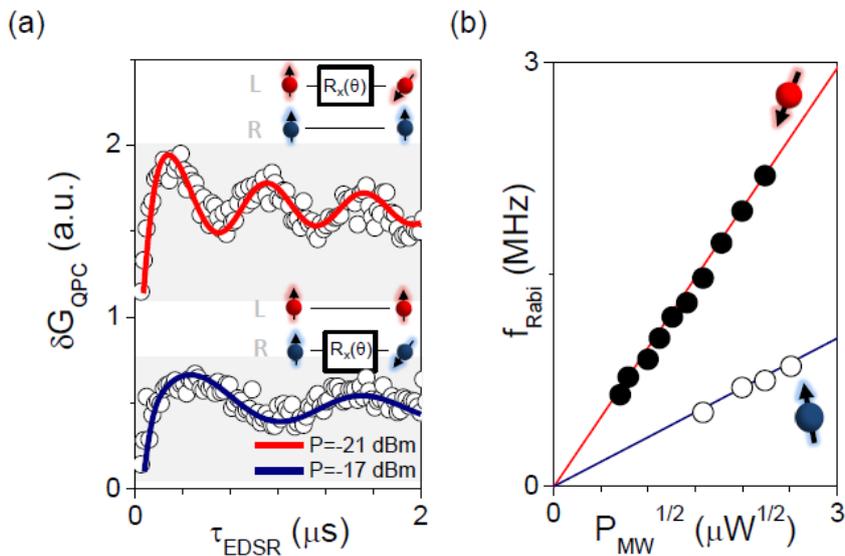

FIG.3 (color)

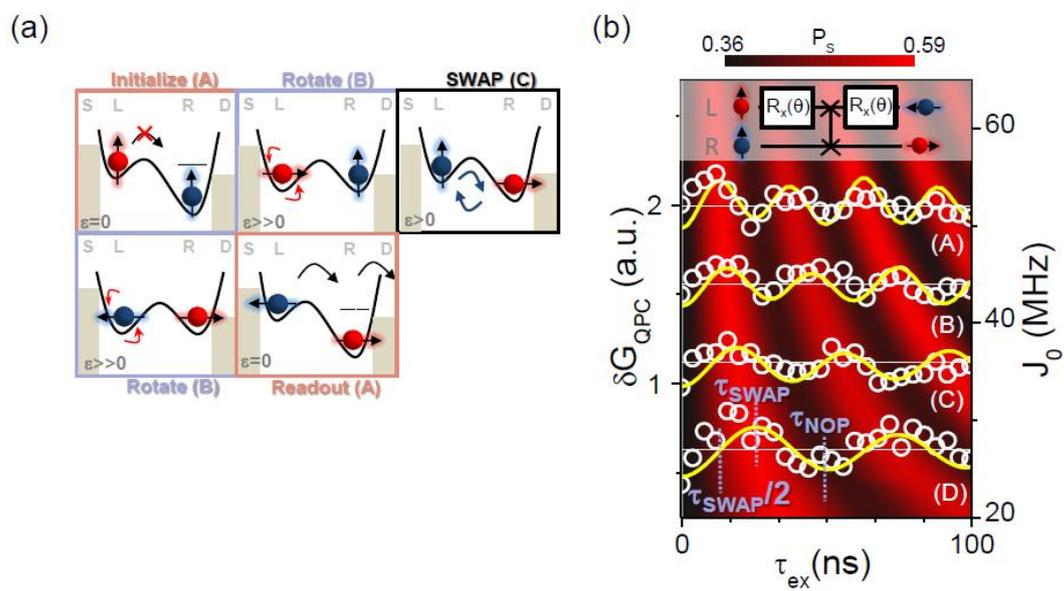

FIG.4 (color)

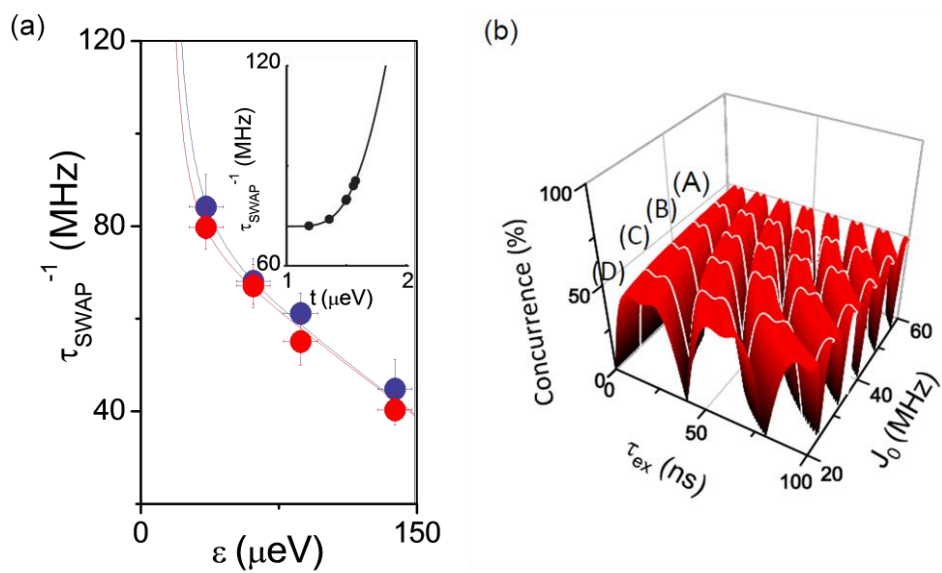